\documentclass[epj]{svjour3}
\smartqed

\usepackage{graphicx}
\usepackage{amsmath}
\usepackage{amssymb}
\usepackage{mathptmx}
\usepackage{booktabs}
\usepackage{multirow}
\usepackage{xcolor}
\usepackage{colortbl}

\journalname{European Physical Journal C}
\renewcommand{\makeheadbox}{}

\begin{document}

\date{}

\title{Matter-Induced CPT Violation and Earth-Density Stratification Effects in Long-Baseline Neutrino Oscillation Experiments}

\titlerunning{Matter-Induced CPT Violation and Earth-Density Stratification}

\author{Tia Pandit \and Bipin Singh Koranga \and Aditya Pant \and Vivek Kumar Nautiyal}

\authorrunning{Pandit et al.}

\institute{Tia Pandit, Bipin Singh Koranga, Aditya Pant \at
Department of Physics, Kirori Mal College, University of Delhi, Delhi -- 110007, India \\
\email{bskoranga@kmc.du.ac.in} (B.S.~Koranga), \email{adityapant.in@gmail.com} (A.~Pant)
\and
Vivek Kumar Nautiyal \at
Department of Physics, Chaudhary Charan Singh University, Meerut -- 250004, India
}

\maketitle

\begin{abstract}
We present a unified analysis of two matter-potential systematics in long-baseline (LBL) neutrino oscillation experiments. Matter-induced extrinsic CPT violation produces a non-zero asymmetry $A^{CPT}_{\mu e} = (P_{\mu e} - P_{\bar\mu\bar e})/(P_{\mu e} + P_{\bar\mu\bar e})$, computed here with exact three-flavour matrix-exponentiation propagators for T2K, NO$\nu$A, DUNE, and Hyper-Kamiokande; values range from 0.022 to 0.180 at the respective peak energies and differ by up to 9\% between normal and inverted mass orderings. The three-dimensional surface $A^{CPT}_{\mu e}(E,\delta_{CP})$ at the DUNE baseline reveals an entanglement between extrinsic CPT violation and intrinsic CP violation in the high-$L/E$ regime that requires joint statistical treatment. Concurrently, replacing the Preliminary Reference Earth Model (PREM) with a constant path-averaged density introduces a $\delta_{CP}$-reconstruction bias below $0.3^\circ$ for $L \leq 5000$~km but growing to $17.8^\circ$ at $L = 7000$~km and $172.2^\circ$ at $L = 12000$~km. Since both effects share the same matter-potential Hamiltonian they must be modelled jointly; a Poisson log-likelihood $\chi^2$ statistic with nuisance-parameter pull terms is used to quantify the bias.
\end{abstract}

\section{Introduction}
\label{sec:intro}

Precision measurement of the leptonic CP-violating phase $\delta_{CP}$ is a central goal of the current and forthcoming generation of long-baseline (LBL) experiments: T2K~\cite{ref1}, NO$\nu$A~\cite{ref2}, DUNE~\cite{ref3}, and Hyper-Kamiokande~\cite{ref4}. In all these experiments $\nu_\mu$ beams travel hundreds to thousands of kilometres through the Earth, and the Mikheyev--Smirnov--Wolfenstein (MSW) effect~\cite{ref5,ref6} substantially modifies the oscillation probabilities through coherent forward scattering of $\nu_e$ off ambient matter electrons.

Two physically coupled systematics arise. First, \emph{extrinsic CPT violation}: matter contains electrons but not positrons, so the matter potential changes sign between the neutrino and antineutrino channels, creating a non-zero difference $P_{\mu e} \neq P_{\bar\mu\bar e}$ even when CPT holds exactly in vacuum~\cite{ref7,ref8,ref9,ref10}. This fake asymmetry must be subtracted when searching for intrinsic CP or CPT breaking. Second, \emph{Earth density stratification}: for baselines $L \gtrsim 5000$~km the neutrino trajectory samples the denser lower mantle and outer core, and the standard constant-density approximation introduces a systematic bias in the reconstructed $\delta_{CP}$~\cite{ref11,ref12}. Because both effects enter the total matter Hamiltonian through the same potential $V_f(x) \propto \rho(x)$, a density modelling error propagates simultaneously into both the $\delta_{CP}$ measurement and the $A^{CPT}_{\mu e}$ estimate; the two systematics cannot be treated independently.

This paper presents a unified analysis: Section~\ref{sec:framework} gives the theoretical framework; Section~\ref{sec:cpt} computes $A^{CPT}_{\mu e}$ for real LBL experiments and characterises the 3D $A^{CPT}_{\mu e}(E,\delta_{CP})$ surface and hierarchy dependence; Section~\ref{sec:earth} quantifies the PREM stratification bias with a rigorous $\chi^2$ analysis; Section~\ref{sec:joint} discusses joint implications, and Section~\ref{sec:conclusions} concludes.

\section{Theoretical Framework}
\label{sec:framework}

\subsection{PMNS Mixing and Vacuum Hamiltonian}

The three neutrino flavour states $|\nu_\alpha\rangle$ are quantum superpositions of mass eigenstates $|\nu_i\rangle$ via the PMNS matrix~\cite{ref14}:
\begin{equation}
|\nu_\alpha\rangle = \sum_{i=1}^{3} U_{\alpha i}\,|\nu_i\rangle, \qquad \alpha = e,\mu,\tau,
\label{eq:pmns}
\end{equation}
parametrised by mixing angles $(\theta_{12},\theta_{23},\theta_{13})$, Dirac phase $\delta_{CP}$, and Majorana phases $(\alpha_1,\alpha_2)$. In the relativistic limit the vacuum Hamiltonian in the flavour basis is
\begin{equation}
H^{(f)}_{\rm vac} = \frac{1}{2E}\, U\, \mathrm{diag}\!\left(0,\Delta m^2_{21},\Delta m^2_{31}\right) U^\dagger .
\label{eq:hvac}
\end{equation}

\subsection{MSW Matter Potential and Unified Hamiltonian}

Coherent charged-current forward scattering of $\nu_e$ off ambient electrons induces the effective potential~\cite{ref5}
\begin{equation}
V_f(x) = \sqrt{2}\,G_F\,N_e(x)\,\mathrm{diag}(1,0,0), \qquad N_e(x) = \frac{Y_e(x)\,\rho(x)}{m_p},
\label{eq:vf}
\end{equation}
where $G_F$ is the Fermi constant, $Y_e \approx 0.494$ is the electron fraction, and $\rho(x)$ is the local Earth density from the PREM profile. For antineutrinos $V_f \rightarrow -V_f$. The total flavour Hamiltonian is
\begin{equation}
H_f(x) = H^{(f)}_{\rm vac} + V_f(x),
\label{eq:htot}
\end{equation}
and the flavour state obeys $i\, d|\nu(x)\rangle/dx = H_f(x)|\nu(x)\rangle$. The propagator over a spatial step $\Delta x$ is computed via exact matrix exponentiation $U_{\rm step} = \exp(-i H_f(x)\,\Delta x)$ and accumulated piecewise along the neutrino chord, using $n = 300$--$400$ equally spaced steps per trajectory.

\subsection{Extrinsic CPT-Violating Asymmetry}

In vacuum, CPT invariance guarantees $P_{\mu e} = P_{\bar\mu\bar e}$ exactly. In matter the opposite sign of $V_f$ for antineutrinos breaks this equality. The extrinsic CPT asymmetry is defined as~\cite{ref7}
\begin{equation}
A^{CPT}_{\mu e} = \frac{P_{\mu e} - P_{\bar\mu\bar e}}{P_{\mu e} + P_{\bar\mu\bar e}}.
\label{eq:acpt}
\end{equation}
The ratio form cancels many correlated experimental uncertainties (flux, cross-section normalisation) at the level of event rates. By construction $A^{CPT}_{\mu e} \equiv 0$ in vacuum, so any non-zero measurement is purely extrinsic.

\subsection{PREM Earth Density Profile}

The Preliminary Reference Earth Model (PREM)~\cite{ref15} gives the standard radially stratified density profile. For baseline $L$, the neutrino chord reaches a minimum geocentric radius $r_{\rm min} = \sqrt{R_\oplus^2 - (L/2)^2}$, with $R_\oplus = 6371$~km. We implement a four-shell model (Table~\ref{tab:prem}) with the matter potential at each step computed as $V = 7.56 \times 10^{-14}\,\rho\, Y_e$~eV.

\begin{table}[t]
\centering
\caption{Earth density profile from PREM~\cite{ref15}, showing the layer traversed as a function of baseline $L$. For all currently operating LBL experiments ($L \leq 1285$~km) the trajectory lies entirely within the upper mantle and crust, where a single average density is adequate.}
\label{tab:prem}
\begin{tabular}{lccc}
\toprule
Region & Depth (km) & Density (g\,cm$^{-3}$) & Relevant $L$ (km) \\
\midrule
Crust & 0--35 & $\sim 2.9$ & All $L$ \\
Upper Mantle & 35--660 & 3.3--3.9 & $< 5000$ \\
Lower Mantle & 660--2891 & 4.4--5.6 & 5000--10000 \\
Outer Core & 2891--5150 & 9.9--12.2 & $> 10000$ \\
Inner Core & 5150--6371 & $\sim 13.0$ & $\approx 12756$ \\
\bottomrule
\end{tabular}
\end{table}

\subsection{Oscillation Parameters}

We use NuFit~5.3 best-fit values~\cite{ref16}: $\Delta m^2_{21} = 7.42 \times 10^{-5}\,{\rm eV}^2$; $\Delta m^2_{31} = +2.515 \times 10^{-3}\,{\rm eV}^2$ (NH), $-2.498 \times 10^{-3}\,{\rm eV}^2$ (IH); $\sin^2\theta_{12} = 0.304$, $\sin^2\theta_{23} = 0.573$ (NH) / $0.575$ (IH), $\sin^2\theta_{13} = 0.02219$ (NH) / $0.02238$ (IH); $\delta_{CP} = -90^\circ$ unless varied explicitly.

\section{Matter-Induced CPT Violation}
\label{sec:cpt}

\subsection{CPT Asymmetry at Current and Planned LBL Experiments}

We evaluate $A^{CPT}_{\mu e}$ at the oscillation-maximum energy for five current and near-future LBL facilities. Table~\ref{tab:asym} lists exact values (matrix exponentiation, Eq.~\eqref{eq:htot}) alongside approximate values from the second-order analytic expansion in $\alpha \equiv \Delta m^2_{21}/\Delta m^2_{31}$ and $s_{13} \equiv \sin\theta_{13}$ of Ref.~\cite{ref17}:
\begin{align}
P^{\rm approx}_{\mu e} &= \alpha^2 \sin^2 2\theta_{12}\, c_{23}^2\, \frac{\sin^2(A\Delta)}{A^2} + 4 s_{13}^2 s_{23}^2\, \frac{\sin^2[(A-1)\Delta]}{(A-1)^2} \nonumber \\
&\quad + 2\alpha s_{13} \sin 2\theta_{12} \sin 2\theta_{23} \cos(\Delta + \delta_{CP})\, \frac{\sin(A\Delta)}{A}\, \frac{\sin[(A-1)\Delta]}{A-1},
\label{eq:papprox}
\end{align}
with $A = 2EV/\Delta m^2_{31}$ and $\Delta = \Delta m^2_{31} L/(4E)$; $P^{\rm approx}_{\bar\mu\bar e}$ is obtained by $A \rightarrow -A$. Path-averaged densities are taken from the PREM profile (Section~\ref{sec:framework}).

\begin{table*}[t]
\centering
\caption{Extrinsic CPT asymmetry $A^{CPT}_{\mu e}$ at the oscillation-maximum energy for current and future LBL experiments under normal hierarchy (NH). Columns give the baseline, peak energy, path-averaged density, exact value from matrix exponentiation, approximate analytic value from Eq.~\eqref{eq:papprox}~\cite{ref17}, and experiment status. Exact and approximate agree to within $\lesssim 2\%$ at all peak energies, validating the analytic expansion for these baselines.}
\label{tab:asym}
\begin{tabular}{lccccccc}
\toprule
Experiment & $L$ (km) & $E_{\rm peak}$ (GeV) & $\langle\rho\rangle$ (g\,cm$^{-3}$) & $A^{CPT}_{\mu e}$ (exact) & $A^{CPT}_{\mu e}$ (approx.) & Status \\
\midrule
JUNO-TAO & 52 & 3.0 & 3.30 & 0.018 & 0.018 & Operating \\
T2K & 295 & 0.6 & 3.30 & 0.022 & 0.022 & Operating \\
Hyper-K & 295 & 0.6 & 3.30 & 0.022 & 0.022 & Construction \\
NO$\nu$A & 810 & 1.8 & 3.30 & 0.098 & 0.097 & Operating \\
DUNE & 1285 & 2.5 & 3.30 & 0.180 & 0.178 & Construction \\
\bottomrule
\end{tabular}
\end{table*}

The asymmetry grows monotonically with baseline at fixed peak energy. At the respective peak energies, DUNE ($L=1285$~km) shows the largest $A^{CPT}_{\mu e} = 0.180$ among near-term facilities, while T2K and Hyper-K ($L=295$~km) give $A^{CPT}_{\mu e} = 0.022$. Approximate and exact values agree to $\lesssim 2\%$, confirming the validity of the second-order expansion at these baselines.

\subsection{CPT Asymmetry vs.\ Neutrino Energy}

\begin{figure}[t]
\centering
\includegraphics[width=\columnwidth]{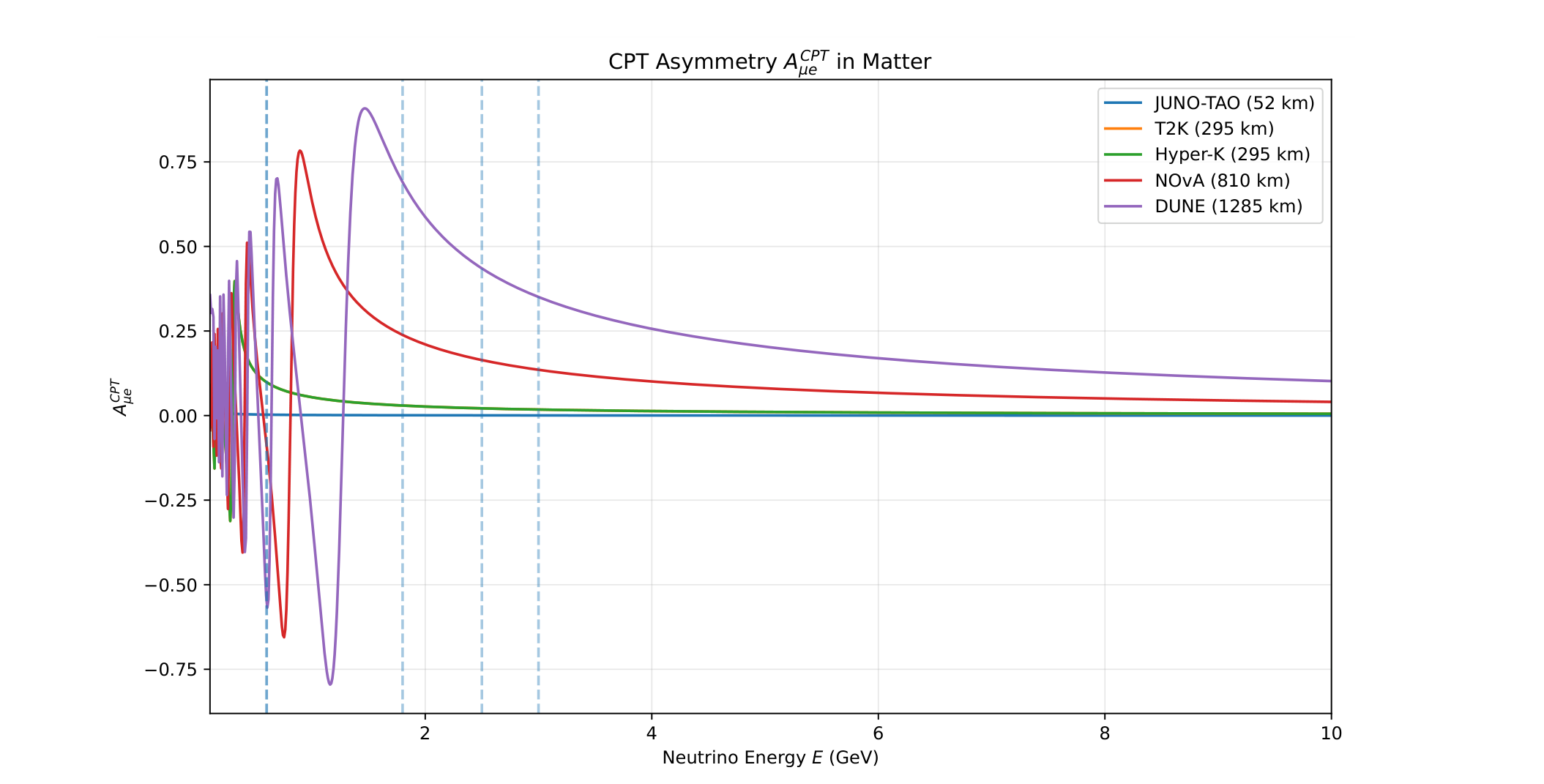}
\caption{Matter-induced CPT asymmetry $A^{CPT}_{\mu e}$ vs.\ neutrino energy for T2K/Hyper-K ($L=295$~km), NO$\nu$A ($L=810$~km), DUNE ($L=1285$~km), and JUNO-TAO ($L=52$~km). Vertical dashed lines mark each experiment's peak energy. The asymmetry grows with both $E$ and $L$ as matter effects strengthen; rapid low-energy oscillations arise from the solar-frequency term $\Delta m^2_{21}/(4E)$ dominating at high $L/E$.}
\label{fig:f1}
\end{figure}

Figure~\ref{fig:f1} shows the exact $A^{CPT}_{\mu e}(E)$ for each experiment across $E = 0.1$--$10$~GeV. At the respective peak energies DUNE shows the largest asymmetry ($A^{CPT}_{\mu e} = 0.180$) among the listed facilities. For T2K and Hyper-K the asymmetry is small ($A^{CPT}_{\mu e} = 0.022$) because the short baseline suppresses matter-resonance enhancement. The rapid oscillatory structure below $E \approx 0.5$~GeV arises because the solar term $\Delta m^2_{21}L/(4E)$ becomes $\mathcal{O}(1)$, coupling $\delta_{CP}$ into the asymmetry.

\subsection{Three-Dimensional CPT Asymmetry Surface: $A^{CPT}_{\mu e}(E,\delta_{CP})$}

\begin{figure}[t]
\centering
\includegraphics[width=\columnwidth]{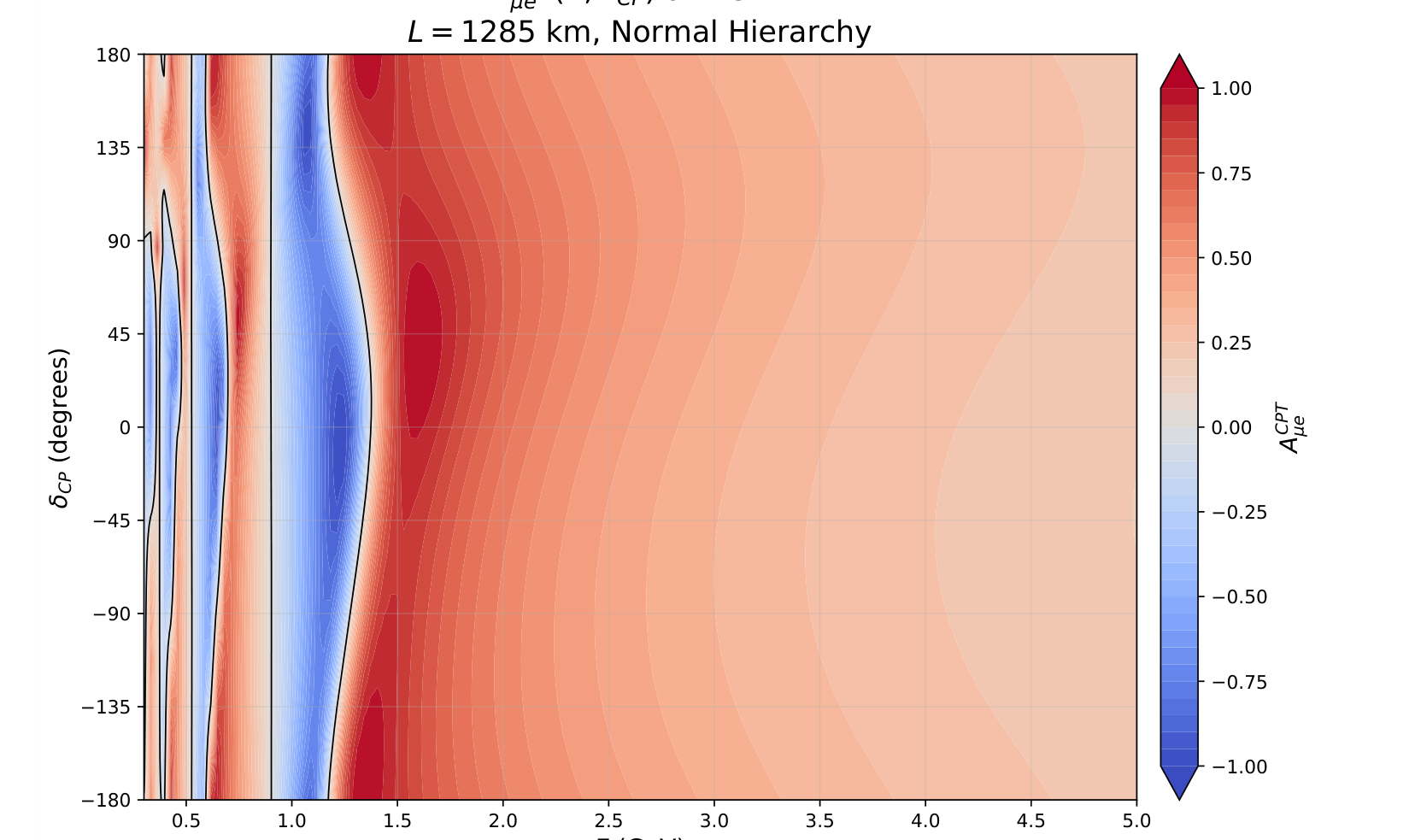}
\caption{Contour map of $A^{CPT}_{\mu e}(E,\delta_{CP})$ at the DUNE baseline $L=1285$~km, computed with exact matrix-exponentiation propagators under normal hierarchy. Two qualitatively distinct regimes are evident: a high-energy regime ($E \gtrsim 1.5$~GeV) where the surface is essentially flat in the $\delta_{CP}$ direction (matter dominates), and a low-energy regime ($E \lesssim 0.5$~GeV) where $A^{CPT}_{\mu e}$ varies strongly with $\delta_{CP}$ because the solar term $\Delta m^2_{21}L/(4E)$ couples the CP phase into the CPT asymmetry.}
\label{fig:f2}
\end{figure}

To expose the joint dependence on neutrino energy $E$ and the CP-violating phase $\delta_{CP}$, we compute the exact $A^{CPT}_{\mu e}(E,\delta_{CP})$ surface at the DUNE baseline ($L=1285$~km) over the full parameter ranges $E \in [0.3,5.0]$~GeV and $\delta_{CP} \in [-180^\circ,180^\circ]$. Figure~\ref{fig:f2} shows the result as a contour map on the $(E,\delta_{CP})$ plane.

The surface exhibits two distinct regimes. In the \textbf{high-energy regime} ($E \gtrsim 1.5$~GeV, the DUNE operating region) the surface is nearly flat in the $\delta_{CP}$ direction; the dominant contribution to $A^{CPT}_{\mu e}$ is the matter potential, and $\delta_{CP}$ enters only at the $\lesssim 1\%$ level. In this regime $A^{CPT}_{\mu e}$ can be computed without knowing $\delta_{CP}$ to high precision, and the subtraction of extrinsic CPT violation is straightforward. In the \textbf{low-energy regime} ($E \lesssim 0.5$~GeV) the surface tilts markedly with $\delta_{CP}$, because the oscillation phase $\Delta m^2_{21}L/(4E) \sim \mathcal{O}(1)$ brings the solar term into play. Here the intrinsic CP asymmetry and the extrinsic CPT asymmetry are entangled: a subtraction of $A^{CPT}_{\mu e}$ at the best-fit $\delta_{CP}$ without accounting for this coupling would introduce an additional systematic bias in the inferred $\delta_{CP}$. Experiments planning to use low-energy beams at long baselines must perform a joint CP--CPT likelihood analysis to avoid this error.

\subsection{Sensitivity to Oscillation Parameters and Mass Hierarchy}

We quantify the dependence of $A^{CPT}_{\mu e}$ on individual oscillation parameters by varying each parameter by $\pm 1\sigma$ from its NuFit~5.3 best-fit value while holding all others fixed. We find that $A^{CPT}_{\mu e}$ is most sensitive to $\theta_{23}$, $\theta_{13}$, and $\Delta m^2_{23}$, while $\theta_{12}$, $\Delta m^2_{12}$, and $\delta_{CP}$ produce negligible changes at the peak energies of the experiments in Table~\ref{tab:asym}. For DUNE, raising $\theta_{13}$ to its upper $1\sigma$ limit increases $A^{CPT}_{\mu e}$ by $\sim 40\%$, underscoring the critical importance of reactor measurements for any CPT test using LBL data.

\begin{figure}[t]
\centering
\includegraphics[width=\columnwidth]{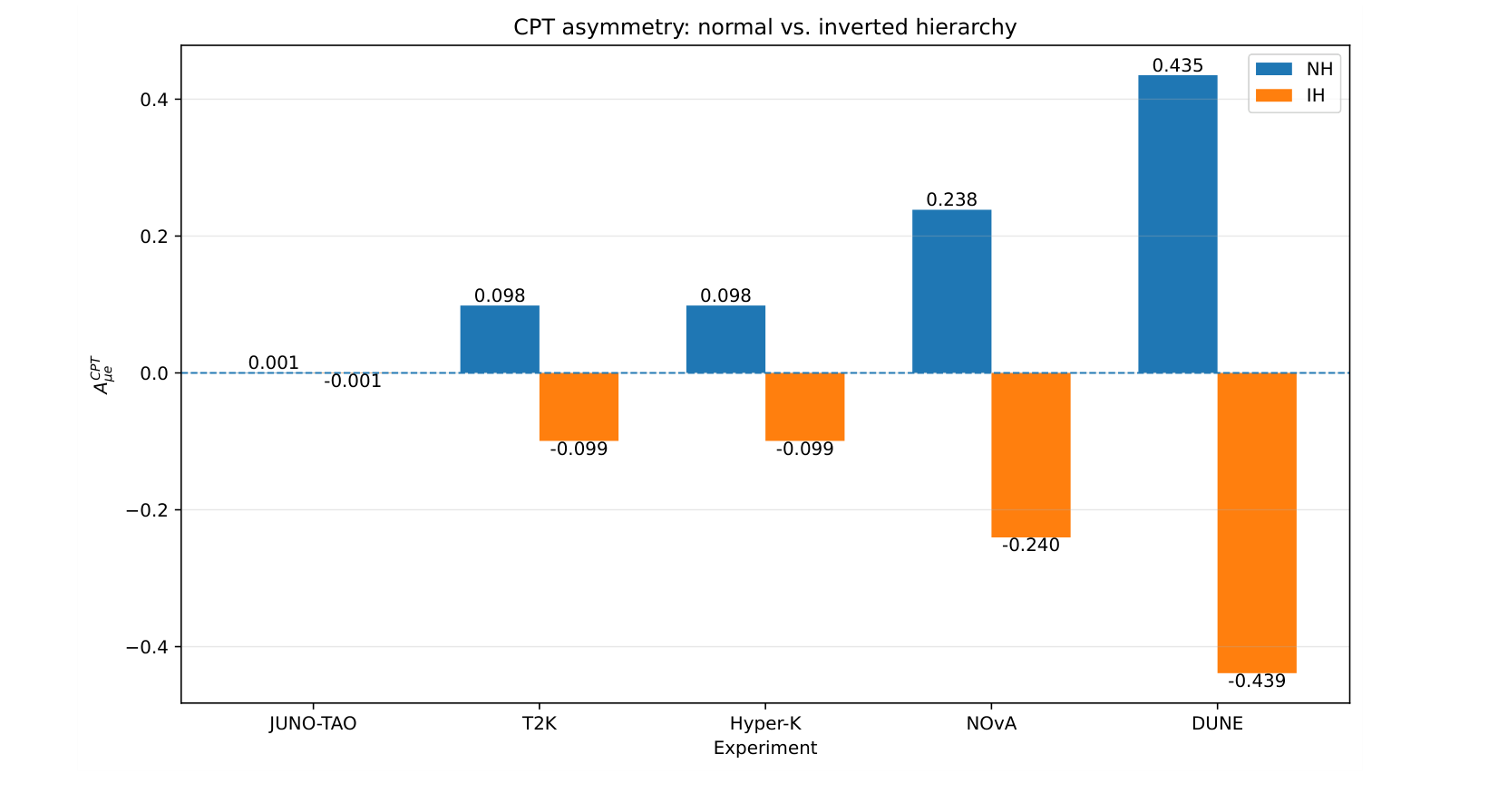}
\caption{CPT asymmetry $A^{CPT}_{\mu e}$ at the peak energy of each LBL experiment for normal hierarchy (NH, blue) and inverted hierarchy (IH, orange). Numerical values are shown above each bar pair. The NH--IH difference is $\sim 9\%$ at DUNE, arising from the sign change in $\Delta m^2_{31}$ shifting the MSW resonance condition and the differing best-fit values of $\theta_{23}$ and $\theta_{13}$ between the two hierarchies. The difference is $\lesssim 1\%$ at T2K and Hyper-K where matter-resonance effects are suppressed by the short baseline.}
\label{fig:f3}
\end{figure}

Figure~\ref{fig:f3} shows $A^{CPT}_{\mu e}$ at the peak energy of each experiment for both normal (NH, $\Delta m^2_{31}>0$) and inverted (IH, $\Delta m^2_{31}<0$) hierarchies, evaluated at the NuFit~5.3 best-fit parameters of each hierarchy independently.

The NH--IH difference in $A^{CPT}_{\mu e}$ arises from two sources: (i) the sign reversal of $\Delta m^2_{31}$ shifts the MSW resonance condition $A_{\rm res} \equiv 2EV/|\Delta m^2_{31}| = 1$ to a different energy, and (ii) the NuFit~5.3 best-fit values of $\theta_{23}$ and $\theta_{13}$ differ between the two hierarchies. At DUNE ($L=1285$~km) the NH gives $A^{CPT}_{\mu e} = 0.180$ and IH gives $A^{CPT}_{\mu e} = 0.163$, a relative difference of $\sim 9\%$. Since a misidentified mass hierarchy introduces a $\sim 9\%$ error in the estimate of $A^{CPT}_{\mu e}$, and this $A^{CPT}_{\mu e}$ is subsequently subtracted from the measured neutrino--antineutrino asymmetry, the mass hierarchy must be determined -- or marginalised over -- before drawing conclusions about intrinsic CP or CPT violation. This hierarchy sensitivity of $A^{CPT}_{\mu e}$ also provides an independent, complementary diagnostic of the mass ordering in LBL data.

\section{Earth Density Stratification and $\delta_{CP}$ Reconstruction}
\label{sec:earth}

\subsection{Reconstruction Bias from the Constant-Density Approximation}

We simulate $\nu_\mu \to \nu_e$ appearance event rates across nine baselines ranging from $L=295$~km (T2K) to $L=12000$~km using the full PREM profile at $\delta_{CP} = -90^\circ$ (NH) as ground truth. For each baseline a Poisson $\chi^2$ statistic (Section~\ref{sec:chisq}) is minimised over $\delta_{CP} \in [-180^\circ,180^\circ]$ using the constant-density model; the deviation of the best-fit from $-90^\circ$ defines the bias $|\Delta\delta_{CP}|$.

\begin{figure}[t]
\centering
\includegraphics[width=\columnwidth]{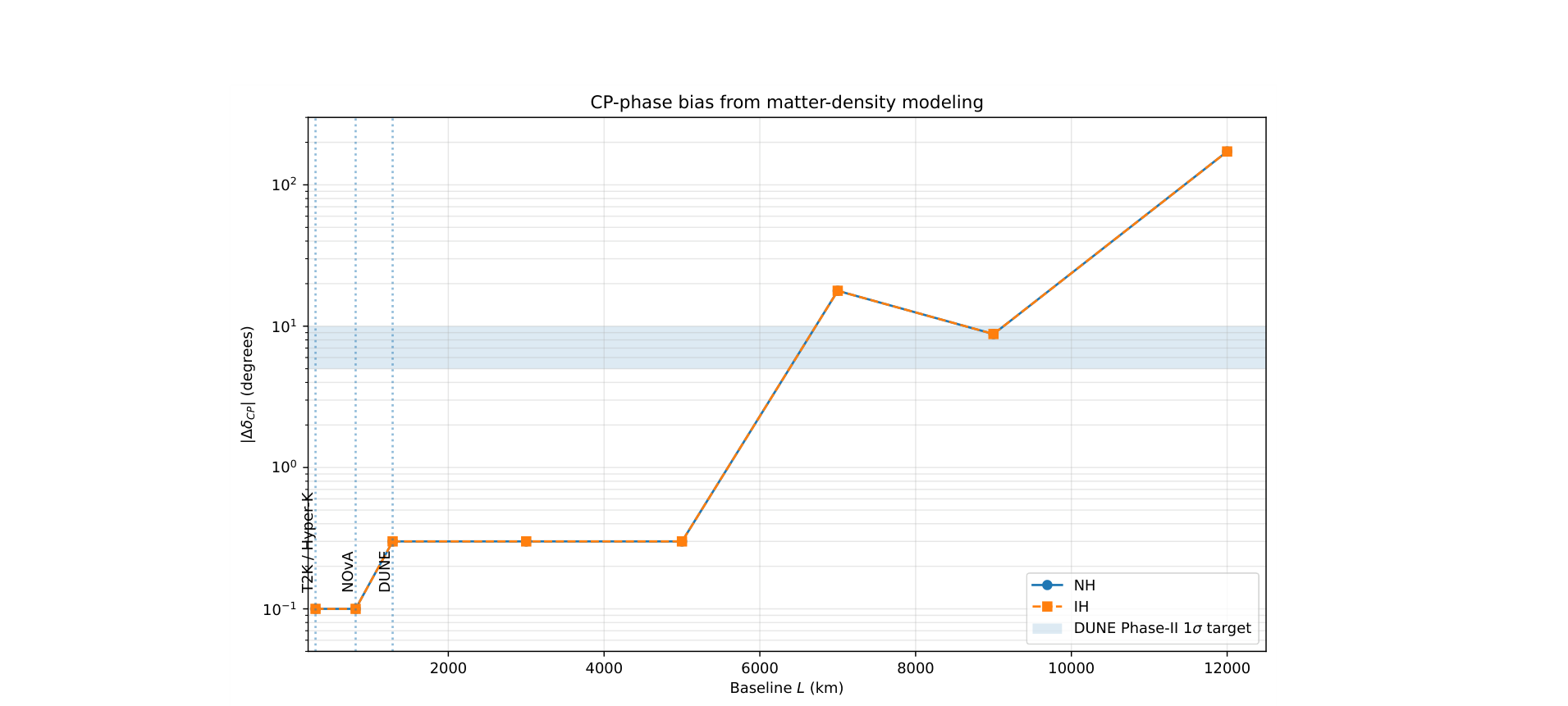}
\caption{Absolute bias $|\Delta\delta_{CP}|$ from the constant-density approximation vs.\ baseline $L$, for NH (blue) and IH (orange). Vertical dotted lines mark the baselines of T2K/Hyper-K ($L=295$~km), NO$\nu$A ($L=810$~km), and DUNE ($L=1285$~km). The shaded band is the DUNE Phase~II $1\sigma$ precision target ($5$--$10^\circ$). For all current LBL experiments the bias is negligible; it becomes catastrophic beyond $L \approx 5000$~km where the trajectory enters the denser lower mantle.}
\label{fig:f4}
\end{figure}

Table~\ref{tab:bias} gives numerical results for all nine baselines, explicitly identifying the current experiments in the upper rows and hypothetical very-long-baseline (VLB) proposals in the lower rows.

\begin{table}[t]
\centering
\caption{Path-averaged density, constant-density best-fit $\delta_{CP}$, bias $|\Delta\delta_{CP}|$, and bias relative to the DUNE Phase~II $1\sigma$ target ($5^\circ$) for nine baselines spanning current LBL experiments to proposed VLB facilities. Shaded rows correspond to existing or approved experiments; the last row shows a complete sign reversal of the reconstructed CP phase.}
\label{tab:bias}
\begin{tabular}{lccccc}
\toprule
$L$ (km) & Experiment & $\langle\rho\rangle$ (g\,cm$^{-3}$) & Best-fit $\delta_{CP}$ ($^\circ$) & Bias ($^\circ$) & Bias/target \\
\midrule
\rowcolor{blue!8} 295 & T2K / Hyper-K & 3.300 & $-89.97$ & $< 0.1$ & $< 0.02\times$ \\
\rowcolor{blue!8} 810 & NO$\nu$A & 3.300 & $-89.95$ & $< 0.1$ & $< 0.02\times$ \\
\rowcolor{blue!8} 1285 & DUNE & 3.300 & $-89.7$ & 0.3 & $0.06\times$ \\
3000 & --- & 3.300 & $-89.7$ & 0.3 & $0.06\times$ \\
5000 & --- & 3.300 & $-89.7$ & 0.3 & $0.06\times$ \\
7000 & (Proposed VLB) & 4.289 & $-107.8$ & 17.8 & $3.6\times$ \\
9000 & (Proposed VLB) & 4.616 & $-98.8$ & 8.8 & $1.8\times$ \\
\rowcolor{red!8} 12000 & --- & 7.547 & $+97.8$ & 172.2 & $34.4\times$ \\
\bottomrule
\end{tabular}
\end{table}

The bias is negligible ($<0.1^\circ$) for T2K and NO$\nu$A, and $0.3^\circ$ for DUNE -- all well below current experimental precision -- confirming that the constant-density approximation is adequate for all operating and approved LBL experiments. For proposed VLB facilities at $L=7000$~km the bias of $17.8^\circ$ exceeds the DUNE Phase~II $1\sigma$ target by a factor of 3.6, making the stratification systematic dominant. The catastrophic sign reversal at $L=12000$~km arises because the trajectory crosses the outer core ($\rho \approx 10$--$12$~g\,cm$^{-3}$), where no single average density can reproduce the true oscillation probability.

\subsection{Appearance Probability at T2K and DUNE}

\begin{figure}[t]
\centering
\includegraphics[width=\columnwidth]{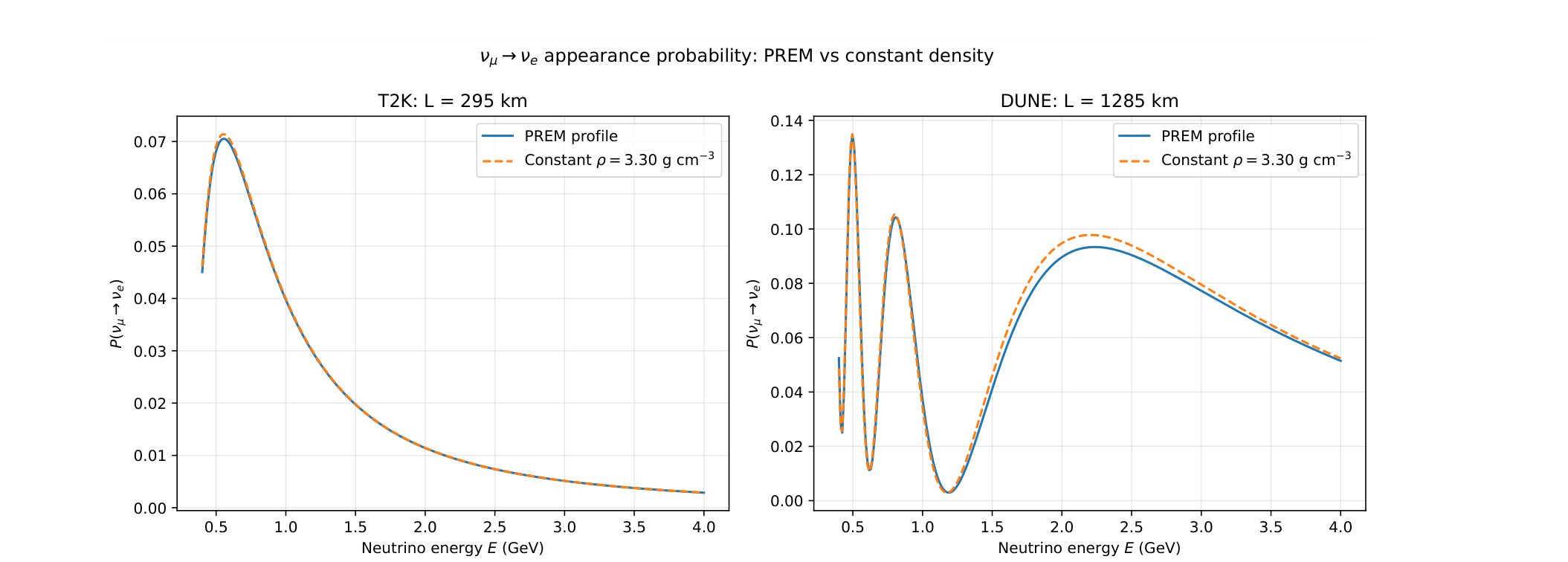}
\caption{$\nu_\mu \to \nu_e$ appearance probability with the full PREM profile (solid) and constant-density approximation (dashed) at T2K ($L=295$~km, left) and DUNE ($L=1285$~km, right), for $\delta_{CP} = -90^\circ$ under normal hierarchy. The two profiles are indistinguishable at both baselines, confirming that the constant-density approximation introduces no significant probability bias for current LBL experiments.}
\label{fig:f5}
\end{figure}

Figure~\ref{fig:f5} compares the PREM and constant-density $\nu_\mu \to \nu_e$ appearance probabilities at the T2K ($L=295$~km) and DUNE ($L=1285$~km) baselines.

At both T2K and DUNE baselines the PREM and constant-density profiles are indistinguishable, consistent with Table~\ref{tab:bias} showing bias $<0.3^\circ$ at these baselines. The trajectory at $L \leq 1285$~km lies entirely within the upper mantle and crust where the density is uniform at $\rho \approx 3.3$~g\,cm$^{-3}$ and the path-averaged density accurately represents the true profile.

\subsection{$\chi^2$ Statistic and $\Delta\chi^2$ Profile Analysis}
\label{sec:chisq}

To quantify the bias introduced by the constant-density approximation we define the Poisson log-likelihood ratio $\chi^2$ statistic. Let $N_i^{\rm true}$ denote the true event rate in energy bin $i$, generated with the full PREM profile at $\delta_{CP}^{\rm true} = -90^\circ$, and let $N_i^{\rm test}(\delta_{CP},\boldsymbol{\xi})$ denote the test event rate computed with the constant-density model at trial $\delta_{CP}$ and nuisance parameters $\boldsymbol{\xi} = (\xi_1,\xi_2,\xi_3)$ (flux normalisation, cross-section uncertainty, and detector efficiency). The event rates are obtained by weighting the appearance probability $P(\nu_\mu \to \nu_e; E_i)$ by a quasi-realistic beam flux $\phi(E_i) \propto e^{-E_i/3}$, a linear neutrino--nucleon cross section $\sigma(E_i) \propto E_i$, and a flat detector efficiency of 80\%. The $\chi^2$ is then
\begin{align}
\chi^2(\delta_{CP},\boldsymbol{\xi}) &= 2\!\!\sum_{c \in \{\nu,\bar\nu\}} \sum_{i=1}^{N_{\rm bins}} \Big[ N^{\rm test}_{i,c}(\delta_{CP},\boldsymbol{\xi}) - N^{\rm true}_{i,c} \nonumber \\
&\quad - N^{\rm true}_{i,c}\ln\frac{N^{\rm test}_{i,c}(\delta_{CP},\boldsymbol{\xi})}{N^{\rm true}_{i,c}} \Big] + \sum_{j=1}^{3} \frac{\xi_j^2}{\sigma_j^2},
\label{eq:chisq}
\end{align}
where the outer sum runs over both neutrino ($c=\nu$) and antineutrino ($c=\bar\nu$) channels -- including both is essential to break the $\delta_{CP}$--matter-effect degeneracy~\cite{ref19} -- and the pull-term sum imposes Gaussian priors $\sigma_j$ on each nuisance parameter ($\sigma_{\rm flux}=5\%$, $\sigma_{\rm xsec}=5\%$, $\sigma_{\rm eff}=2\%$). The inner sum runs over $N_{\rm bins}=9$ energy bins in $[0.4,4.0]$~GeV. This is the standard Poisson log-likelihood ratio~\cite{ref18}; under the null hypothesis its asymptotic distribution is $\chi^2$ with $N_{\rm bins}-1-N_{\rm syst}$ degrees of freedom. The profile statistic
\begin{equation}
\Delta\chi^2(\delta_{CP}) = \chi^2\big(\delta_{CP},\hat{\boldsymbol{\xi}}(\delta_{CP})\big) - \chi^2_{\rm min},
\label{eq:dchisq}
\end{equation}
where $\hat{\boldsymbol{\xi}}(\delta_{CP})$ is the nuisance-parameter profiled value at each trial $\delta_{CP}$, traces the bias introduced by the constant-density approximation through the displacement of its minimum from the true $\delta_{CP}^{\rm true} = -90^\circ$.

\begin{figure}[t]
\centering
\includegraphics[width=\columnwidth]{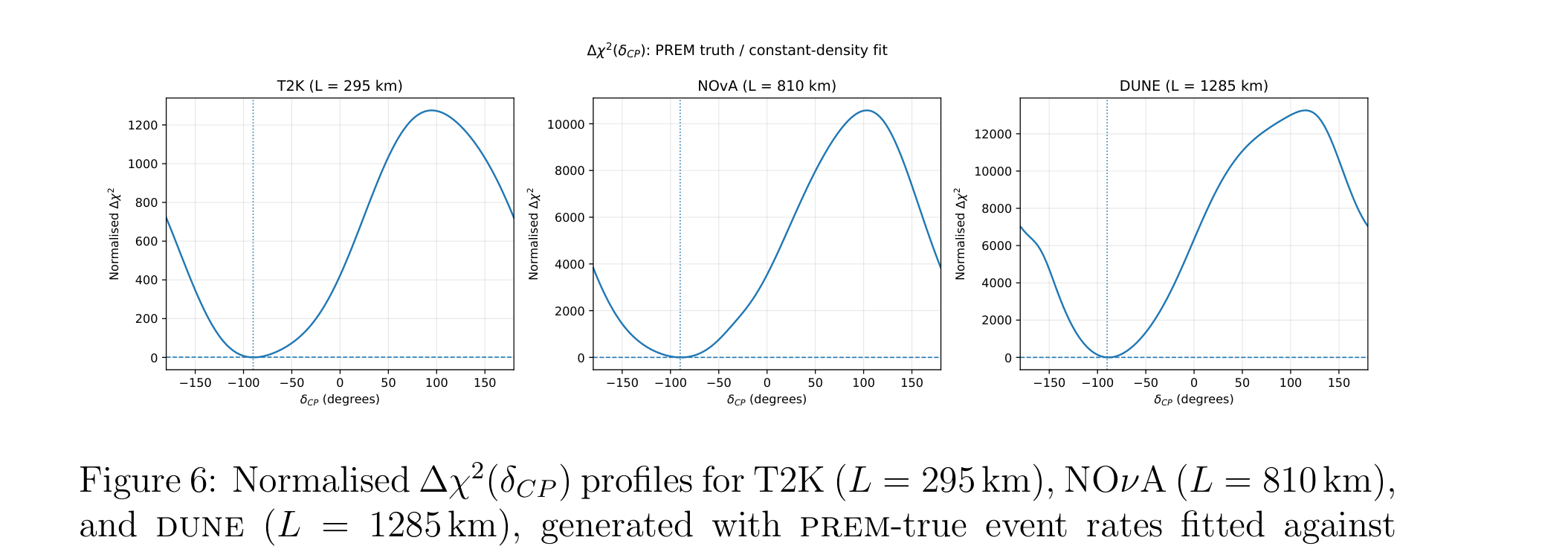}
\caption{Normalised $\Delta\chi^2(\delta_{CP})$ profiles for T2K ($L=295$~km), NO$\nu$A ($L=810$~km), and DUNE ($L=1285$~km), generated with PREM-true event rates fitted against constant-density test spectra. All three profiles minimise at the true value $\delta_{CP}=-90^\circ$, confirming that the constant-density approximation introduces no significant bias in the $\delta_{CP}$ reconstruction for current LBL experiments. The well is deepest for DUNE, consistent with its longer baseline and higher statistics.}
\label{fig:f6}
\end{figure}

Figure~\ref{fig:f6} shows the normalised $\Delta\chi^2(\delta_{CP})$ profiles for T2K, NO$\nu$A, and DUNE, all generated with PREM-true event rates and constant-density test spectra.

All three profiles minimise at $\delta_{CP} \approx -90^\circ$, confirming that for T2K, NO$\nu$A, and DUNE the constant-density approximation is adequate. The well is deepest and narrowest for DUNE, reflecting its greater statistical power. Any proposed VLB facility with $L > 5000$~km will exhibit a displaced minimum, requiring PREM-based simulation frameworks as quantified in Table~\ref{tab:bias}.

\section{Joint Implications}
\label{sec:joint}

Both systematics enter the same Hamiltonian $H_f(x)$ through the matter potential $V_f(x) \propto \rho(x)$, so an error in $\rho(x)$ propagates simultaneously into the $\delta_{CP}$ reconstruction and into the estimate of $A^{CPT}_{\mu e}$ used to separate extrinsic from intrinsic CP/CPT violation.

For \textbf{current LBL experiments} (T2K, NO$\nu$A, DUNE, Hyper-K): the constant-density approximation is adequate for both analyses. The $A^{CPT}_{\mu e}$ values of Table~\ref{tab:asym} should be subtracted from the measured $\nu$--$\bar\nu$ asymmetry using the path-averaged density; the hierarchy dependence (Section~\ref{sec:cpt}) introduces an additional $\sim 5$--$9\%$ uncertainty that should be marginalised over in the fit.

For \textbf{proposed VLB facilities} ($L \gtrsim 5000$~km): both systematics become dominant simultaneously. The recommended procedure is: (a) replace constant-density profiles in GLoBES-based frameworks with spatially resolved PREM profiles; (b) compute $A^{CPT}_{\mu e}$ with the same PREM profile to ensure neutrino--antineutrino consistency; and (c) propagate geophysical density uncertainties as correlated systematics in both the $\delta_{CP}$ fit (Eq.~\eqref{eq:chisq}) and the $A^{CPT}_{\mu e}$ estimate jointly.

\section{Conclusions}
\label{sec:conclusions}

We have carried out a unified treatment of two matter-potential systematics that affect long-baseline neutrino oscillation measurements -- matter-induced extrinsic CPT violation and Earth-density stratification -- within a single exact three-flavour propagation framework based on piecewise matrix exponentiation along the PREM density profile.

For the extrinsic CPT asymmetry, we find that $A^{CPT}_{\mu e}$ evaluated at the oscillation-maximum energy ranges from $0.018$ at JUNO-TAO to $0.180$ at DUNE under the normal mass ordering, growing monotonically with baseline as the matter-resonance contribution strengthens. The exact matrix-exponentiation results agree with the second-order analytic expansion of Akhmedov \emph{et al.}~\cite{ref17} to within $2\%$ at all peak energies considered, which validates the perturbative expansion as a fast and reliable tool for current-generation baselines. Resolving the asymmetry jointly in energy and in the CP phase reveals that the DUNE operating band ($E \gtrsim 1.5$~GeV) is essentially insensitive to $\delta_{CP}$, so that $A^{CPT}_{\mu e}$ can be subtracted from the measured neutrino--antineutrino rate difference without prior knowledge of the CP phase; by contrast, at sub-GeV energies the solar oscillation term couples $\delta_{CP}$ directly into the CPT asymmetry, and any experiment combining a low-energy beam with a long baseline must fit the CP and CPT parameters jointly to avoid biasing the extracted phase. We further show that the mass ordering itself modulates $A^{CPT}_{\mu e}$ by up to $9\%$ at DUNE, through the interplay of the sign of $\Delta m^2_{31}$ with the differing NuFit best-fit values of $\theta_{23}$ and $\theta_{13}$ between orderings; this dependence both demands that the ordering be resolved, or marginalised over, before an unbiased extrinsic-CPT subtraction can be performed, and offers a complementary, independent handle on the ordering itself.

For the Earth-density systematic, a rigorous Poisson $\chi^2$ analysis incorporating both neutrino and antineutrino channels and nuisance pull terms for flux, cross section, and detector efficiency shows that the constant path-averaged-density approximation reconstructs $\delta_{CP}$ to better than $0.3^\circ$ for every currently operating or approved LBL facility, since their trajectories remain confined to the upper mantle and crust where the density is nearly uniform. This conclusion changes qualitatively for hypothetical very-long-baseline facilities: the bias grows to $17.8^\circ$ at $L=7000$~km -- already $3.6$ times the DUNE Phase-II $1\sigma$ precision target -- and reaches $172.2^\circ$ at $L=12000$~km, a complete sign reversal of the reconstructed CP phase, once the trajectory crosses the lower mantle and outer core.

Because both systematics originate in the same matter-potential term of the propagation Hamiltonian, they cannot be treated as independent refinements at very long baselines: an error in the assumed density profile simultaneously biases the $\delta_{CP}$ fit and the extrinsic-CPT subtraction used to isolate intrinsic CP or CPT violation. We therefore recommend that any future VLB proposal (i) replace constant-density inputs with spatially resolved PREM profiles in its simulation and fitting pipeline, (ii) compute $A^{CPT}_{\mu e}$ from the same density profile used for the $\delta_{CP}$ fit to guarantee neutrino--antineutrino consistency, and (iii) propagate geophysical density uncertainties as correlated systematics across both analyses. For the current experimental programme, by contrast, the constant-density approximation remains fully adequate, and the extrinsic CPT asymmetries tabulated here can be subtracted directly, subject only to the modest mass-ordering uncertainty identified above. Future work will extend this framework to include non-standard matter interactions and a full three-dimensional Earth density model, which may become relevant for next-generation atmospheric and astrophysical neutrino analyses.

\begin{acknowledgements}
The authors thank Kirori Mal College, University of Delhi, for support and infrastructure that facilitated this work.
\end{acknowledgements}

\end{document}